# Spin glass behavior in amorphous CrSiTe$_3$ alloy

Xiaozhe Wang[1], Jiayue Wang[1], Yihui Jiang[1], Suyang Sun[1,2,*], Jiangjing Wang[1,*], Riccardo Mazzarello[3], Wei Zhang[1,*]

[1]Center for Alloy Innovation and Design (CAID), State Key Laboratory for Mechanical Behavior of Materials, Xi'an Jiaotong University, Xi'an, China.
[2]Institute of Materials, Henan Academy of Sciences, Zhengzhou 450046, China.
[3]Department of Physics, Sapienza University of Rome, Rome 00185, Italy

Emails: sy.sun@hnas.ac.cn, j.wang@xjtu.edu.cn, wzhang0@mail.xjtu.edu.cn

**Abstract**
Owing to the intrinsically high crystallization temperatures, layered phase-change materials, such as CrGeTe$_3$ and InGeTe$_3$, are attracting attention for embedded memory applications, In addition to the electrical contrast, a major change in magnetic properties is observed in CrGeTe$_3$ upon switching from the crystalline to the amorphous state. In this work, we report a combined *ab initio* modeling and magnetic characterization study on the isostructural silicon parent compound of CrGeTe$_3$, namely, CrSiTe$_3$. Amorphous CrSiTe$_3$ has similar structural properties to amorphous CrGeTe$_3$; however, it shows a smaller energy difference between the ferromagnetic configuration and the random magnetic configuration, indicating a high probability of spin glass formation. Indeed, direct-current and alternating-current magnetic measurements show that the coercive force of amorphous CrSiTe$_3$ is higher than that of amorphous CrGeTe$_3$. Therefore, the pinning effect of spins is enhanced in amorphous CrSiTe$_3$, leading to a more robust spin glass state with a higher freezing temperature. The large magnetic contrast between the amorphous and crystalline phase could make CrSiTe$_3$ a potential candidate for phase-change magnetic switching applications.

Chalcogenide phase-change materials (PCMs) are one of the leading candidates for nonvolatile memory and neuromorphic in-memory computing applications.[1-8] The basic principle is to utilize the large contrast in electrical or optical properties between the amorphous state and the crystalline state of PCMs for memory encoding. The flagship $Ge_2Sb_2Te_5$ (GST) alloy has been employed in memory devices for several years. By introducing impurities or excess Ge into GST, its crystallization temperature $T_x$ can be enhanced from 150 °C to 260 °C and above for embedded memory applications.[9-13] In GST, crystallization proceeds via the formation of three-dimensional (3D) seeds that consists of cubic rocksalt-like structures with octahedral bonds.[14-18] As an alternative to the extrinsic alloying of conventional PCMs, new types of two-dimensional (2D) layered PCMs [19-26] with intrinsically high crystallization temperatures above ~260 °C have been discovered, including $CrGeTe_3$ (CrGT) and $InGeTe_3$ (InGT). The structural origin of the large difference in $T_x$ between amorphous GST and amorphous CrGT/InGT is attributed to the formation of 2D-like seeds in the latter.[24-26] Such 2D-like seeds are structurally complex, since they consist of mixed tetrahedral and octahedral bonds in a trilayer structural pattern.[24]

Alloying GST with $3d$ transition metal impurities (~7 at%), including Cr, Mn and Fe, opens up the possibility of fast magnetic switching upon reversible phase transition between the amorphous and the crystalline states.[27-29] In both states, the magnetic coupling is ferromagnetic (FM), and the average magnetic moment can be enhanced by up to ~30% upon crystallization. The smaller magnetic moment in the amorphous phase is attributed to the more compact bonding configuration, which induces stronger saturation effects.[29] Regarding CrGT, the concentration of magnetic atoms is much higher, i.e., 20 at%, and the crystalline state shows a clear FM order with a Curie temperature $T_C$ of ~61 K.[30, 31] In Ref. [25], we made amorphous CrGT powder samples and identified a clear spin glass state with a freezing temperature $T_f$ ~8 K in the presence of external magnetic fields. Later, S. Zhang *et al.* prepared amorphous CrGT bulk samples via high-energy heavy-ion irradiation and they showed the sample to be ferromagnetic with a much higher $T_C$ of ~200 K compared to the crystalline state.[32] They attributed the different magnetic order to the defective local structures induced by strong irradiation. The FM behavior with high $T_C$ was also observed in amorphous Fe doped GST.[27]

In this work, we investigate the magnetic interactions in the amorphous phase of the isostructural silicon parent compound of CrGT, namely, $CrSiTe_3$ (CrST).[30, 33-36] Crystalline (c-) CrST takes the same atomic structure as c-CrGT, but shows a strongly anisotropic 2D Ising-like FM order with a smaller $T_C$ of ~32 K.[30] The different magnetic interactions due to the substitution of Ge with Si atoms could also lead to changes in the magnetic properties of the amorphous phase. According to our previous structural and electrical measurements,[24] the thermal stability of amorphous (a-) CrST is enhanced with respect to a-CrGT, and thin films crystallize at $T_x$ ~300 °C. The c-CrST thin films show higher electrical resistance values as compared to the amorphous phase, similarly to CrGT. However, it remains elusive whether amorphous a-CrST forms a FM state or a spin glass state at low temperatures and whether the $T_C$ value or the $T_f$ value increases or decreases with respect to CrGT. To address these questions, we carry out *ab initio* simulations and magnetic measurements to gain an in-depth understanding of the magnetic interactions in a-CrST.

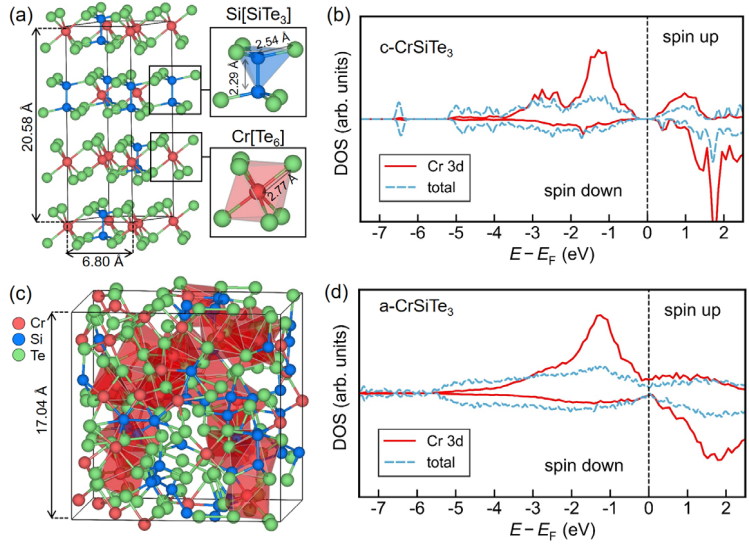

**Figure 1. (a)-(b)** The DFT-relaxed atomic structure of c-CrST in the FM state and the corresponding total and projected DOS. **(c)-(d)** The AIMD-generated melt-quenched a-CrST model in the FM state and the corresponding total and projected DOS.

We carried out density functional theory (DFT) and DFT-based *ab initio* molecular dynamics (AIMD) melt-quenched simulations[37-39] to model the crystalline and amorphous phase of CrST, following our previous work.[24] The atomic structures were relaxed at 0 K, prior to the electronic structure and magnetic calculations using the VASP code,[40] for which the projector augmented wave (PAW) method,[41] the Perdew–Burke–Ernzerhof (PBE) functional[42] and the semi-empirical DFT-D3 van der Waals corrections[43] were used. Fig. 1a shows the DFT-relaxed atomic structure of c-CrST. Locally, a Cr atom forms six octahedral bonds with Te atoms, and two Si atoms form tetrahedral bonds with six atoms. The two Si-centered tetrahedral units share one Si–Si homopolar bond. The bond length of Cr–Te, Si–Te and Si–Si bonds is 2.77 Å, 2.54 Å and 2.29 Å, respectively. In total, the c-CrST model contains 6 Ge, 6 Si and 18 Te atoms in a hexagonal unit cell with $a_{hex}$ = 6.80 Å and $c_{hex}$ = 20.58 Å. We computed the electronic structure of the FM ground state and showed the total density of states (DOS) together with the projected DOS on the Cr atoms in Fig. 1b. The majority spin states of the Cr 3$d$ orbitals lie below the Fermi level $E_F$, and the minority spin states are almost empty and located at much higher energies. It is known that the FM order in c-CrST is mainly stabilized by the Cr–Te–Cr (bond angle ~90°) superexchange mechanism.[44] Nevertheless, both c-CrST and c-CrGT have a narrow band gap (0.52 eV vs 0.34 eV), suggesting that carrier-mediated *p*–*d* exchange interactions may also help stabilize the FM order.

Regarding the amorphous phase, we calculated three melt-quenched amorphous models via independent thermal processes, and one such model is displayed in Fig. 1c. The a-CrST model contains 36 Ge, 36 Si and 108 Te atoms in a cubic cell with a lattice edge of 17.04 Å. The local structural order around Cr and Si atoms are mostly preserved in the amorphous phase. The structural details of a-CrST can be found in our previous work.[24] Here, we focus on the electronic structure and magnetic properties. Fig. 1d shows the DOS and projected DOS of one a-CrST model in the FM state. The energy gap is closed in the amorphous phase, providing more carriers for transport, which results in a higher conductivity than the crystalline state. This behavior is consistent with our electrical measurements.[24] The overall distribution of occupied and unoccupied states is similar to the crystalline counterpart, but the peaks are in general broader in the amorphous phase. We repeated the electronic structure calculations for the other two a-CrST models, which gave very similar results.

Obviously, it is not feasible to construct precise antiferromagnetic (AFM) configurations in the amorphous phase similar to the ones found in the layered crystalline phase (e.g. the Néel, zigzag, and stripy AFM configurations) due to the disordered distribution of Cr atoms. Instead, we computed random magnetic (RM) configurations by flipping half of the moments of the Cr atoms in the FM configuration in a random fashion. For a given a-CrST FM model (Fig. 2a), three RM configurations were created and their atomic coordinates were further relaxed (Fig. 2 b-d). Atomic relaxation resulted in some small atomic displacements from the FM structure. The average total energy of the three RM states is 1.07, 2.85 and 2.93 meV/atom higher than the FM state. The above procedure was repeated for the two a-CrST models. In total, we obtained three FM states and nine RM states, and the average energy difference between RM and FM a-CrST is ~2.82 meV/atom. This value is lower than that of a-CrGT, ~4 meV/atom.[26] The smaller energy difference indicates that a-CrST may tend to form a spin glass state more easily than a-CrGT. Fig. 2e shows the distribution of local moments of Cr atoms in the three a-CrST FM states. The respective maximum, minimum and average moment is 3.57, 2.11 and 3.02 $\mu_B$ for a-CrST, in comparison with that of c-CrST, 3.15 $\mu_B$. Fig. 2f shows the distribution of local moments in nine a-CrST RM states. The spin up and spin down moments show a slightly different distribution, but the average net magnetic moment is close to zero, ~0.01 $\mu_B$. The randomly assigned negative moments also resulted in some difference in the distribution of Cr–Te bonds, as presented in Fig. 2g and 2h.

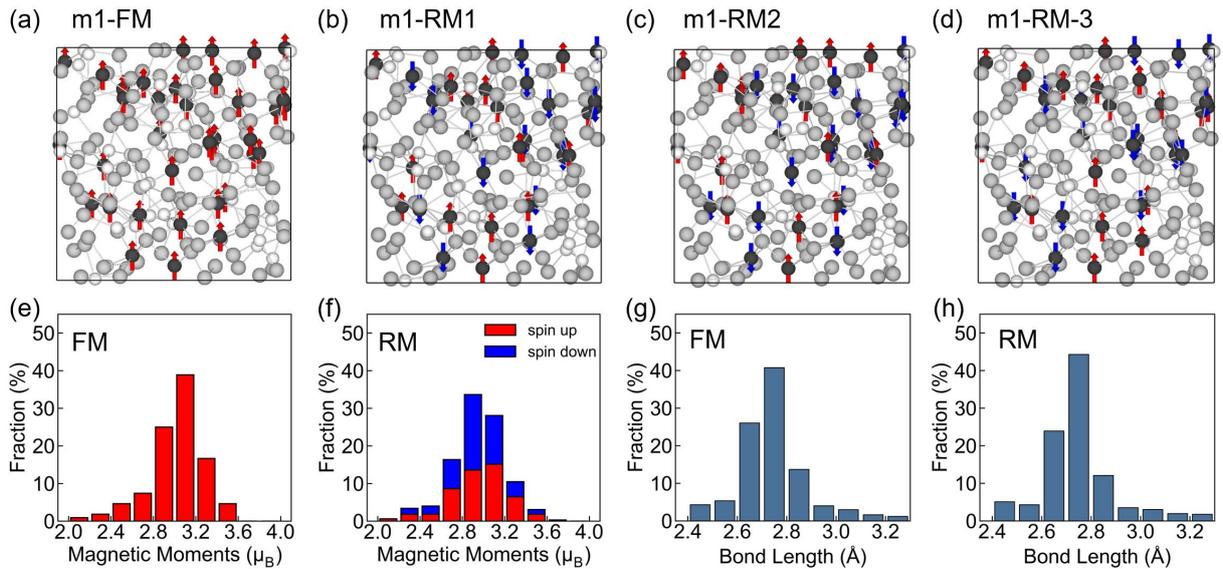

**Figure 2. (a)-(d)** The FM state and three RM states of an a-CrST model. The spin up and spin down moments are indicated by the red and blue arrows. **(e)-(h)** The distribution of magnetic moments and Cr–Te bonds in the a-CrST models. The data were collected over three FM states and nine RM states.

In the following, we discuss our experimental results for a-CrST. Our magnetic measurements confirmed our prediction of spin glass behavior in a-CrST. We followed the synthesis procedure used for a-CrGT powder sample for a-CrST as well, namely, we deposited multiple a-CrST thin films on oil-based ink coated $SiO_2$/Si substrates via magnetron co-sputtering (AJA Orion-8) of three high-purity Cr, Si and Te elemental targets, immersed the as-deposited thin films in acetone to exfoliate them from the substrate, and finally obtained the powder sample after the cleaning and drying process. The total weight of the CrST powder sample was ~4.8 mg, which was sufficient to enable detectable signals for magnetic measurements using our superconducting quantum interference device (SQUID) magnetometer setup (Quantum Design MPMS3-VSM). We performed transmission electron microscopy (TEM) measurements (Talos-F200X operated at 200 kV) and confirmed that the CrST powder sample formed the amorphous phase, as no visible crystallites were observed in bright-field image and the corresponding selected area electron diffraction (SEAD) pattern showed dim halos (Fig. 3a inset).

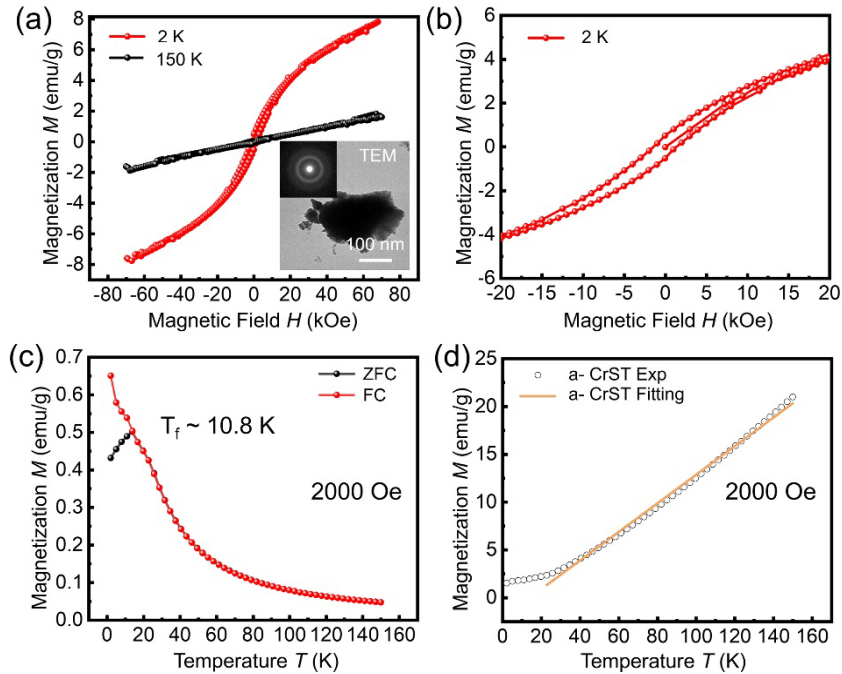

**Figure 3. (a)** The *M-H* curve measured at 2 K and 150 K with a maximum DC field of 70 kOe. Inset: the TEM image and the SAED pattern of the a-CrST powder sample. **(b)** The zoomed-in *M-H* curve measured at 2 K. **(c)** The ZFC and FC *M-T* curves measured under an applied field of 2000 Oe. **(d)** The linear fitting of the inverse susceptibility $\chi^{-1}(T)$ under 2000 Oe in the FC condition by the Curie-Weiss law.

First, we measured the magnetic response of the amorphous (a-) CrST powder sample under direct-current (DC) magnetic fields. The *M-H* curve exhibits a linear characteristic at 150 K, corresponding to the paramagnetic state, but shows a clear hysteresis loop at 2 K (Fig. 3a). Fig. 3b shows a zoom-in version of the *M-H* curve at 2K. As the external magnetic field decreases gradually from 70 kOe to zero, the magnetization of a-CrST does not drop to zero but retains a certain residual magnetization, 0.4991 emu/g, which is slightly larger than that of a-CrGT 0.1822 emu/g. The hysteresis loop indicates the presence of FM order. When the DC magnetic field strength exceeds 10 kOe, the *M-H* curve shows a linear shape until the maximum magnetic field is reached, 70 kOe. Yet, the magnetization of a-CrST is not saturated at 70 kOe, suggesting the existence of collinear AFM moments or randomly distributed magnetic moments. The coercivity of a-CrST, 1.952 kOe (nearly zero in c-CrST[35]), is larger than that of a-CrGT, 1.15 kOe at 2 K, which means that a-CrST is more resistant to external magnetic interferences than a-CrGT.

We measured the change of magnetization with temperature under a given external magnetic field. Fig. 3c shows the *M-T* curve under 2000 Oe. A clear splitting in magnetization between the zero-field-cooled (ZFC) and field-cooled (FC) curves can be observed as the temperature was reduced to ~10.8 K, revealing a clear spin glass feature. Under the ZFC condition, the spin directions are randomly frozen in the absence of external magnetic field, resulting in a lower susceptibility. In contrast, the applied magnetic field under the FC condition guides spin ordering during cooling, leading to a higher susceptibility. As *T* was further increased to 40–80 K, a rapid decay in magnetization was observed, and the whole *M-T* curve became flat above 140 K, indicating the transition to the paramagnetic state. By fitting the *M-T* curve above 20 K in the FC mode using the Curie-Weiss law $\chi^{-1} = (T - \theta)/c$, we obtained a positive value, $\theta$ = 13.83 K, in contrast with that of a-CrGT, $\theta$ = −15.22 K. [25]

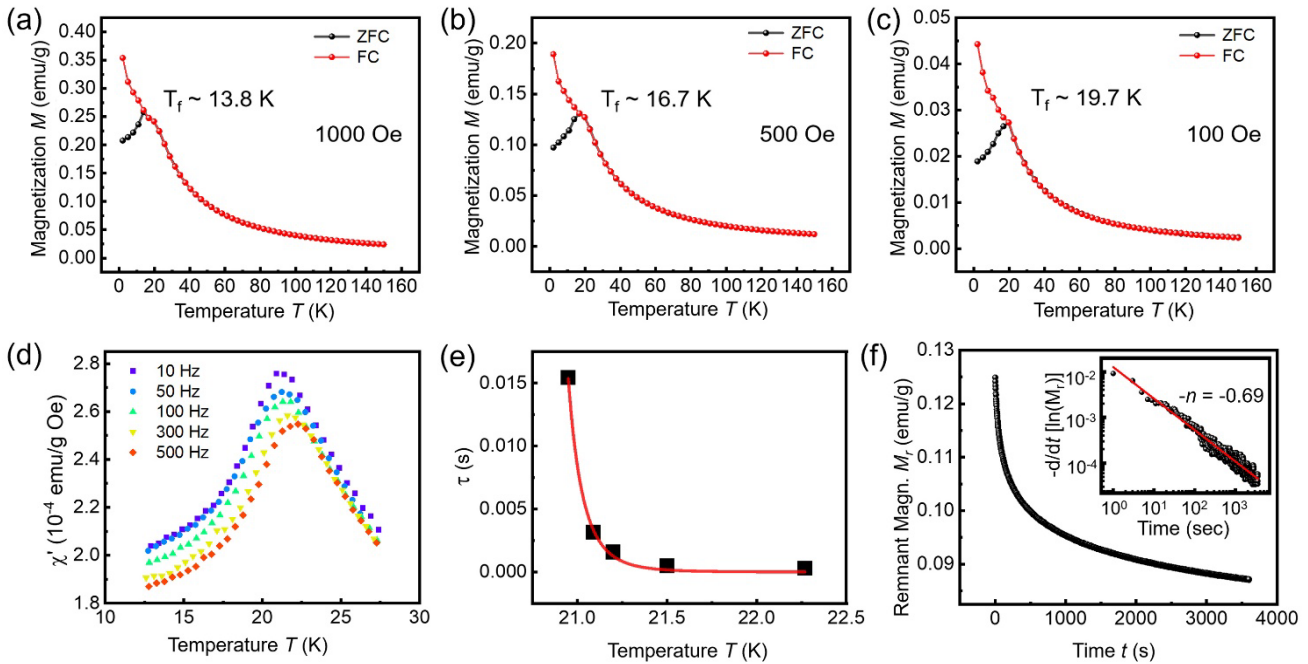

**Figure 4. (a)-(c)** The ZFC and FC *M-T* curves measured under an applied field of 1000 Oe, 500 Oe and 100 Oe. **(d)** The temperature dependence of the real part of the AC susceptibility measured at frequencies *f* =10, 50, 100, 300, and 500 Hz with an AC magnetic field of 7.5 Oe. **(e)** The fitting of the relaxation time against the power law. **(f)** Time dependence of the isothermal remnant magnetization $M_r$ after removal of an initial DC magnetic field of 10 kOe at *T* = 2 K. Inset: fitting of the change of $M_r$ with time on a logarithmic scale.

We repeated the temperature-dependent magnetic measurements under different magnetic fields. As displayed in Fig. 4a-c, all the recorded ZFC and FC curves showed a clear splitting at low temperatures. As the strength of the external field decreased, the $T_f$ value was gradually increased, reaching ~19.7 K under 100 Oe. This trend is consistent with the typical spin glass behavior. In comparison with a-CrGT, a-CrST forms a stronger spin glass state. Under 2000 Oe, a-CrGT showed no splitting between the ZFC and FZ modes. A $T_f$ value of ~7 K was obtained as the field was reduced to 1000 Oe, which increased to ~10.9 K under 100 Oe applied filed. For a given applied field, a-CrST showed a higher $T_f$, than a-CrGT, which can be attributed to the fact that a-CrST required more energy to overcome the greater coercive forces in order to induce depinning of the spins.

We carried out alternating-current (AC) magnetic measurements to gain further understanding of the spin glass state of a-CrST. The applied AC magnetic field was 7.5 Oe, and the measured frequency ranged from 10 Hz to 500 Hz. As shown in Fig. 4d, the peak of the AC magnetic susceptibility $\chi'$ (real part) decreases with increasing frequency, and the peak temperature $T_p$ increases from 20.9 K to 22.3 K. The following equation can be used to quantitatively determine the spin glass behavior of a-CrST: $\Phi = \Delta T_p/(T_p \Delta log_{10} f)$, where $T_p$ is the peak temperature and $f$ is the frequency. The obtained $\Phi$ is 0.0503 for a-CrST, which falls in the typical range of canonical spin glass systems, 0.0045–0.08.[45] The relaxation time $\tau$ of a spin glass near the transition temperature can be expressed by $\tau = \tau_0 [T_p/T_0 - 1]^{-zv}$, where $T_0$ is the finite static freezing temperature, $\tau_0$ is the characteristic flipping time of the magnetic moments, and $zv$ is the dynamical critical exponent. Normally, $zv$ is found between 4 and 13, and $\tau_0$ equals ~$10^{-10}$–$10^{-13}$ s.[45] According to the fitting results, our a-CrST powder sample shows $T_0$ = 20.47 K, $zv$ = 5.94 and $\tau_0$ = 3.05×$10^{-12}$ s (Fig. 4e). The higher $T_0$ compared to the one of a-CrGT ($T_0$ = 8.43 K) indicates that a-CrST forms a more robust spin glass state.

At last, we performed time-dependent remanent magnetization $M_r$ measurements. After cooling the a-CrST sample down to 2 K in the ZFC mode, we applied a DC magnetic field of 10 kOe for 5 minutes. Then we removed the magnetic field, and monitored the evolution of the isothermal remnant magnetization with time. As shown in Fig. 4f, a slow decay was observed. By plotting the time dependent magnetization in logarithmic scale as log {-d/dt [lnM$_r$]} and log {t}, we can obtain a linear fitting of the curve. The slop –$n$ equals –0.69 for a-CrST (that of a-CrGT is –0.66), which is consistent with the typical spin glass behavior.

In conclusion, we have thoroughly investigated the magnetic properties of amorphous CrST. Our DFT calculations yield a smaller energy difference between the ferromagnetic and random magnetic configurations in a-CrST than in a-CrGT, suggesting that a-CrST should also form a spin glass state at low temperatures. The DC and AC magnetic measurements confirmed that a-CrST is a spin glass with a higher freezing temperature than a-CrGT. The coercive force in a-CrST was measured to be higher than that in a-CrGT, which implies that more energy is required to overcome the depinning of spins, leading to a more robust spin glass state in a-CrST. The large magnetic contrast between the ferromagnetic crystalline phase and the spin glass amorphous phase makes CrST a potential candidate for phase-change magnetic switching applications.

**Acknowledgment**
W.Z. thanks the support of National Natural Science Foundation of China (62374131). J.-J.W. thanks the support of National Natural Science Foundation of China (62204201). The authors thank Qizhong Zhao for his technical support on magnetic measurement. We acknowledge the HPC platform of Xi'an Jiaotong University and Computing Center in Xi'an for providing computational resources. The International Joint Laboratory for Micro/Nano Manufacturing and Measurement Technologies of XJTU is acknowledged. R.M. gratefully acknowledges funding from the PRIN 2020 project "Neuromorphic devices based on chalcogenide heterostructures" funded by the Italian Ministry for University and Research (MUR).

**Competing interests**
The authors declare no competing interests.

**Data Availability Statement**
Data supporting this work are available upon reasonable requests to the corresponding authors.